# CSI-aided MAC with Multiuser Diversity for Cognitive Radio Networks

Yuan Lu, Alexandra Duel-Hallen
Department of Electrical and Computer Engineering,
North Carolina State University
Raleigh, NC, 27606
{ylu8, sasha}@ncsu.edu

*Abstract*—**Cognitive Radio (CR) aims to increase the spectrum utilization by allowing secondary users (SU) to access unused licensed spectrum bands. To maximize the throughput given limited sensing capability, SUs need to strike a balance between sensing the channels that are not heavily used by primary users (PU) and avoiding collisions with other SUs. To randomize sensing decisions without resorting to multiuser sensing policies, it is proposed to exploit the spatially-variant fading channel conditions on different links by adapting the reward to the channel state information (CSI). Moreover, the proposed channel-adaptive policy favors links with high achievable transmission rate and thus further improves the network throughput.**

## I. INTRODUCTION

We consider an overlay CR structure [1], where SUs can access the spectrum if there are no active PUs in the neighborhood and thus must sense the spectrum prior to access. Each SU makes its own sensing decision without coordination of a central controller. Given limited sensing capability and constrained resources [2], an efficient sensing strategy involves a tradeoff between sensing the channels that are not likely to be occupied by PUs and diversifying sensing decisions of different SUs to avoid potential secondary collisions (i.e., collisions among SUs).

In single user approaches to sense and access, SUs do not contend for the right to sense the spectrum and make sensing decisions according to the expected rewards. For example, in [3], the PU traffic is modeled as a slotted Markov process, and the internal state of this process is unknown at each time slot. Each SU employs the belief vector that reflects its knowledge of the current channel state to make sensing decisions. As neighboring SUs accumulate the statistical information about the PU traffic, they are likely to converge to the same set of channels, resulting in congestion and increased secondary collision rate. In this case, the actual reward can be much smaller than the expected reward, leading to low throughput.

*Related Work*: References [4-6] address the SU congestion in a CR network by randomizing the sensing decisions of SUs. A static CR network is considered in [4], where the probability to sense each channel is assigned to every SU, and the sensing policy is formulated as an optimization problem over all combinations of these probabilities. This sensing policy assumes that SUs are willing to sacrifice individual throughput to boost the network throughput, which is not always a practical assumption. To avoid convergence to the same channels, [5] proposes a simple randomized sensing policy where the channel selection probability of each SU is given by the normalized belief probability. In [6], the cognitive medium access is proposed for uncertain environments where the PU traffic statistics are unknown a priori and have to be learned and tracked. In the single user scenario, the multiarmed bandit problem is formulated to obtain the tradeoff between exploring statistical information of PU traffic (long-term gain) and sensing the channels with the highest estimated availability probability given current available information (short-term gain). Multiuser scenarios are also considered where the channel selection approach is formulated as an optimization problem for cooperative SUs and a non-cooperative game for selfish SUs, respectively.

Negotiation among SUs has been also proposed to avoid SU congestion. A negotiation policy is investigated in [7] for a two-user two-channel case. However, this approach is not a practical solution for CR network with more than two SUs since the control overhead may grow exponentially as the number of SUs increases. Also, the order of negotiation needs to be determined carefully, or the negotiation messages will collide. A negotiation-based sensing policy is developed in [8]. Without taking PU traffic into consideration, it aims to achieve the maximum spectral coverage by spreading sensing decisions of SUs over different channels. To conclude, strategies in the literature resolve SU congestion by compromising their ability to sense their favorite channel.

*Our Contribution*: Most previous work on sensing strategies focuses PU traffic and ignores the impact of channel conditions. Moreover, all policies in the literature employ a reward proportional to the channel bandwidth. This reward is the same for all SU pairs. With this reward choice, neighboring SUs converge to the same channels when single user sensing policies are employed, so multiuser approaches are necessary to randomize sensing decisions. We propose to exploit spatial variation of SU links to resolve competition among SUs and to improve the throughput. In particular, we adapt the reward to the CSI (i.e. the channel gain) of individual SU pairs. This CSI varies over the locations of SUs due to path loss and fading, thus providing multiuser diversity. Using multipath and shadow fading channel models, we demonstrate that this reward choice improves the throughput and randomizes sensing decisions without resorting to multiuser sensing approaches.

---

This research was supported by the ARO grant W911NF-10-1-0394 and NSF grant CNS-1018447)

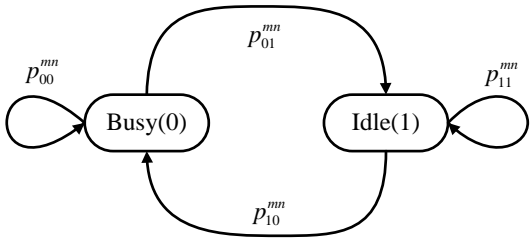

Fig 1: State Transition of PU traffic for channel $n$ at the $m^{th}$ SU location

The rest of this paper is organized as follows. In section II, we present the system model and describe the proposed CSI-aided MAC protocol. The numerical results are provided in section III. Conclusions and future work are discussed in section IV.

## II. CSI-AIDED MAC PROTOCOL

We consider a CR network with $M$ SUs and $N$ non-overlapping channels. The PU traffic is modeled as a stationary Markov process with independently evolving channels. The transition probabilities are assumed known to all SUs. For channel $n$ at the $m^{th}$ SU location, $p_{ij}^{mn}$ denotes the probability of transition from state $i$ to state $j$, where $i, j \in \{0(busy), 1(idle)\}$ as shown in Fig 1. Both the primary network and the CR network share the same slotted structure and all users (including PUs and SUs) are perfectly synchronized. Each SU can sense only one channel at a given time slot. We assume the transmission range of PUs is much larger than that of SUs. In this case, it is very likely that the SU transmitter and the SU receiver are exposed to the same spectrum opportunities, which can be detected at the SU transmitter side only [3]. However, it is still beneficial for the SU receiver and other SUs who does not have data to transmit (or receive) to sense the channels periodically and thereby collecting the PU traffic statistics for future use.

### A. Belief Update

At time slot $t$, the belief vector is denoted as $\boldsymbol{\theta}^m(t) = [\theta^{m1}(t), ..., \theta^{mn}(t), ..., \theta^{mN}(t)]$, where $\theta^{mn}(t)$ is the conditional probability that channel $n$ is available at the $m^{th}$ SU location based on past sensing outcomes. The initialization and the update process of the belief vector for the myopic policy [7] is described in TABLE I. for the $m^{th}$ SU.

### B. CSI-Aided Myopic Sensing Policy

The objective of a sensing policy is to maximize the expected reward given the belief vector. In the literature, the reward is given by the channel bandwidth B, which is often normalized to one for all channels. This reward choice leads to a simple myopic, or greedy, policy in with $R^{mn}(t) = B$, where the channel that is the most likely to be idle in the current time slot is sensed. (The myopic policy can be easily extended to the case when channel bandwidths are different.) This policy has good performance [3] and is optimal when all channels have the same transition probabilities [9] in the single SU pair scenario. However, this policy ignores competing SUs, so its performance is degraded when multiple SUs are active. Since the transmission range of PUs is typically much larger than that of SUs, the neighboring SUs are very likely to be affected by the same set of PUs, and the belief vectors $\boldsymbol{\theta}^m(t)$ will converge to similar values for these SUs as $t$ increases. As a result, neighboring SUs will tend to choose the same channel to sense, but at most one SU can transmit successfully on that channel, even with a very sophisticated collision-avoidance scheme. Other SUs will have to back off, leading to reduced individual throughput. Moreover, SUs will compete for the same channels, leaving other channels unexploited and thus degrading the network throughput.

TABLE I. BELIEF INITIATION AND UPDATE PROCESS

| Step | Description |
|---|---|
| 1 | At the first time slot $t = 1$, the initial belief vector is given by the stationary probabilities of the Markov process: $\boldsymbol{\theta}^m(1) = [\theta^{m1}(1), ..., \theta^{mn}(1), ..., \theta^{mN}(1)]$ $= [\frac{p_{01}^{m1}}{p_{01}^{m1} + p_{10}^{m1}}, ..., \frac{p_{01}^{mn}}{p_{01}^{mn} + p_{10}^{mn}}, ..., \frac{p_{01}^{mN}}{p_{01}^{mN} + p_{10}^{mN}}]$ |
| 2 | At each time slot $t$, SU $m$ chooses to sense the channel $n_*^m(t)$ by maximizing the expected reward $R^{mn}(t)$: $n_*^m(t) = \arg\max_n \theta^{mn}(t) R^{mn}(t)$ Assuming no sensing errors, the sensing result $a^m(t) = 1$ if the channel is idle and $a^m(t) = 0$ otherwise. |
| 3 | The belief vector is updated according to the sensing result $a^m(t)$, $\theta^{mn}(t+1) = \begin{cases} p_{11}^{mn}, & \text{if } n_*^m(t) = n, a^m(t) = 1 \\ p_{01}^{mn}, & \text{if } n_*^m(t) = n, a^m(t) = 0 \\ \theta^{mn}(t) p_{11}^{mn} + (1 - \theta^{mn}(t)) p_{01}^{mn}, & \text{if } n_*^m(t) \neq n \end{cases}$ |
| 4 | Step 2 and Step 3 is repeated over the time horizon $t \in [1, T]$. |

To address this issue, a randomized sensing policy is required. When the reward is the same for all SUs, it is necessary to modify the myopic policy to diversify the sensing decisions. However, in a wireless CR network, it is beneficial to relate the reward to the channel strength in addition to the bandwidth. Since the powers of the SU links differ greatly due to the path loss and channel fading, this reward choice randomizes the sensing decisions. Moreover, this choice results in improved expected transmission rate. In summary, we propose to adapt to the random variation of wireless channels observed at different SU links to diversify channel sensing decisions and to improve the CR network throughput.

Consider the $m^{th}$ SU transmitter-receiver pair. At time slot $t$, the received signal-to-noise ratio (SNR) of the link between these two SUs on the $n^{th}$ channel is given by $\gamma^{mn}(t) = g^{mn}(t) P / N_0$, where $g^{mn}(t)$ is the received channel gain (the CSI), $P$ is the transmission power, and $N_0$ is the power spectral density of the complex Additive White Gaussian noise. We assume the value of $g^{mn}(t)$ (or, equivalently, the value of $\gamma^{mn}(t)$) is known at the SU transmitter $m$ and is fixed for the duration of the slot. If

channel $n$ is chosen by the $m^{th}$ SU pair, the expected channel capacity is

$$E[C^{mn}(t)] = \theta^{mn}(t) \cdot B_n \log_2(1 + \gamma^{mn}(t)/B_n) \quad (1)$$

where $B_n$ is the bandwidth of the $n^{th}$ channel. In the proposed CSI-aided myopic sensing strategy, SU $m$ chooses to sense the channel with the maximum expected channel capacity,

$$n_*^m(t) = \arg\max_n \{E[C^{mn}(t)]\} \quad (2)$$

Note that the reward given by the channel capacity

$$R^{mn}(t) = C^{mn}(t) \quad (3)$$

will differ among SU pairs, randomizing the sensing decisions and improving the individual throughput. The proposed sensing method adapts to the CSI at the SU transmitter. While in this paper, the reward is given by the maximum achievable rate, adaptive modulation and other forms of adaptive transmission can be employed in practice. Moreover, we assume perfect CSI at the SU transmitter, but in practice the CSI needs to be estimated and predicted [10], and the effect of interference should also be taken into account in computing the CSI.

The SU will transmit over the channel if it is sensed to be idle or go to sleep during the current time slot if the channel is busy. If multiple SU pairs choose to sense the same channel, and the channel is idle, we assume that only one of them can transmit successfully. This can be accomplished by using a carefully designed medium access (MAC) scheme, which will be discussed in detail in the next section.

### C. Medium Access Control

In the previous section, we have discussed how different SU pairs should make sensing decisions with the aid of the CSI. It is also important to discuss the MAC decisions for each SU pair to guarantee that the SU transmitter and the SU receiver switch to the same channel for data transmission. Moreover, to avoid secondary collisions, some level of negotiation among different SU pairs is also required. We assume that there is one out-of-band control channel, over which SUs can exchange control packets, and that each SU has a database recording the status of all its neighbors. The slot structure is illustrated in Fig 2. In this paper, we use a slightly modified mutli-channel MAC protocol of [11,12], which is described below. Denote the transmitter and the receiver of the $m^{th}$ SU link as $TX_m$ and $RX_m$, respectively.

1) At the beginning of each time slot $t$, $TX_m$ chooses a channel to sense according to the proposed sensing policy (Please refer to Section II-B).

2) Suppose the $n^{th}$ channel is chosen. If this channel is free, $TX_m$ will continue to the next step; otherwise, it goes to sleep until the next time slot.

3) $TX_m$ checks its local database to see if there are any active SU receivers in the neighborhood. If not, it will send a short request-to-send (RTS) packet to $RX_m$ on the control channel. The spectrum opportunity is assumed to be symmetric, so the $n^{th}$ channel should be available at $RX_m$ as well.

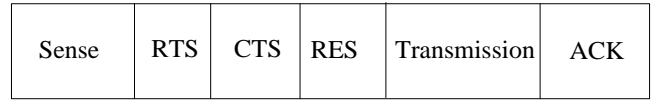

Fig 2: The Slot Structure

4) On receiving the RTS, $RX_m$ checks its local database to see if there are any active SU transmitters in the neighborhood. If not, it will send a short clear-to-send (CTS) packet back to $TX_m$ on the control channel. Then, $RX_m$ will switch to the $n^{th}$ channel immediately and get ready for data reception.

5) On receiving the CTS, $TX_m$ will broadcast a reservation (RES) packet to inform its neighboring SUs on the control channel. Then it will switch to the $n^{th}$ channel and begin transmitting data to $RX_m$.

6) On overhearing the CTS, the neighbors of $RX_m$ will update their local databases and refrain from transmitting on the $n^{th}$ channel.

7) On receiving the RES, the neighbors of $TX_m$ will update their local databases and refrain from receiving on the $n^{th}$ channel.

8) At the end of each time slot, all SU pairs will stop transmission or wake up and go to step 1.

To avoid collisions, the SUs should listen to the control channel for a predefined amount of time before sending any control packets. If the control channel is quiet during this period, the SU can send a control packet after a random backoff time; otherwise, it will clear its counter and wait for another predefined amount of time. In this paper, we assume the control packets are so short that they never collide.

## III. NUMERICAL RESULTS

We assume the CR network traffic is backlogged initially and remains backlogged over the entire time horizon. To illustrate the potential advantages of the CSI-aided sensing strategy, we consider two fading models: the independent and identically distributed (i.i.d) short-term Rayleigh fading model and the correlated long-term log-normal shadow fading model where short-term fading is removed using diversity techniques. We consider the two distributions separately in our simulations as in [13] for the following reasons. At lower speeds, the log-normal shadowing remains almost constant for the duration of one time slot, and the Rayleigh fading changes sufficiently slowly to allow prediction and adaptation at the SU transmitter. As the speed increases, prediction of the Rayleigh fading component might become infeasible, especially over the wide spectrum of CR systems [10], but it is still beneficial to adapt to the shadow fading [13]. Moreover, shadow fading does not vary significantly with frequency for each SU link in most propagation scenarios [14]. Thus, estimation and tracking of shadow fading CSI is simpler and more practical than for short-term fading CSI.

### A. I.I.D Rayleigh Fading Model

First, we employ a CR network with 20 SU pairs and with 40 channels with the same bandwidth $B=1$ evolving independently. The transition matrix of the channel state is the

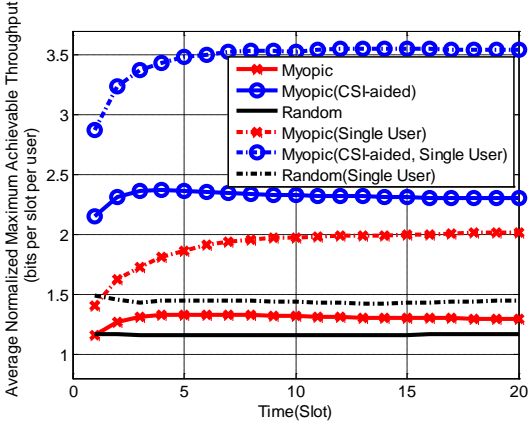

Fig 3: Comparison of CSI-aided and conventional sensing strategies; 20 SU pairs, 40 channels; i.i.d Rayleigh fading channel; average SNR=10dB.

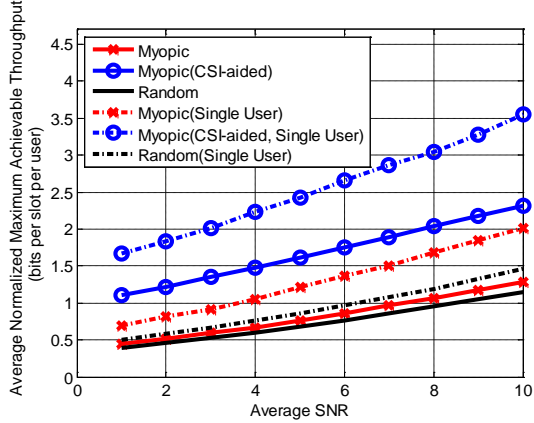

Fig 4: Throughput of sensing strategies vs average SNR; 20 SU pairs, 40 channels; i.i.d Rayleigh fading channel.

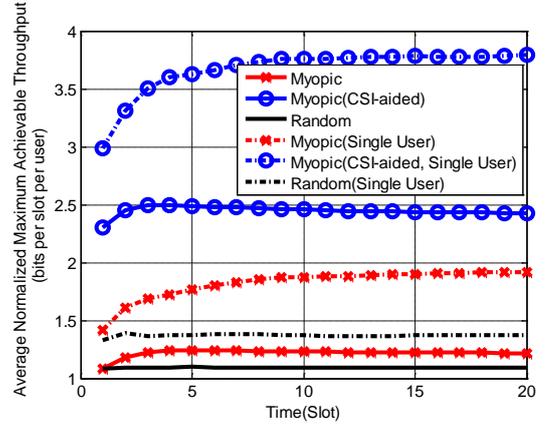

Fig 5: Comparison of CSI-aided and conventional sensing strategies; 20 SU pairs, 40 channels; correlated log-normal fading; average SNR=10dB; $\rho = 0.2$; $\sigma_{\gamma_{dB}} = 5dB$.

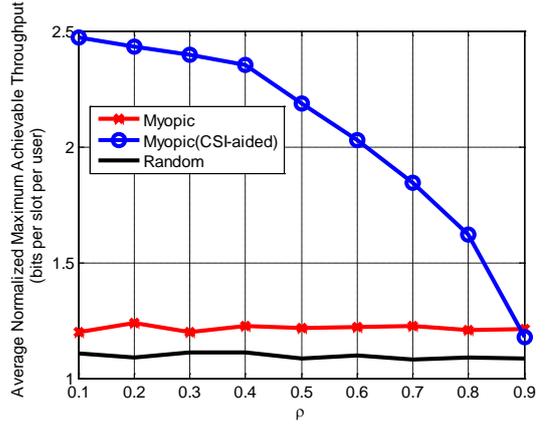

Fig 6: Throughput for different correlation coefficients; 20 SU pairs, 40 channels; correlated log-normal fading; average SNR=10dB; $\sigma_{\gamma_{dB}} = 5dB$

same for all channels and all SUs,

$$P = \begin{bmatrix} p_{00} & p_{01} \\ p_{10} & p_{11} \end{bmatrix} = \begin{bmatrix} 0.8 & 0.2 \\ 0.2 & 0.8 \end{bmatrix} \quad (4)$$

Suppose all SU links are subject to independent Rayleigh fading with the same pdf

$$f(\gamma) = \frac{1}{\mu_\gamma} e^{-\gamma/\mu_\gamma}, \quad (5)$$

where $\mu_\gamma = E[\gamma]$ is the average SNR, chosen to be 10 dB here. We assume very low mobility. Thus, the instantaneous SNR $\gamma$ is fixed over the duration of the transmission (20 slots) and assumed to be known at the transmitter for each channel although in practice it is slowly varying and needs to be estimated and predicted.

In Fig 3, the average normalized maximum achievable throughput is compared for multiuser and single user scenarios for three sensing strategies: the random channel selection approach, the conventional myopic policy that does not adapt to the CSI when making sensing decision, and the proposed approach, i.e., myopic (CSI-aided) strategy. The average normalized maximum achievable throughput is the average throughput obtained using channel capacity as the accumulated reward (i.e., adaptive transmission is assumed once the channel is sensed) divided by the number of SU pairs in the network.

Note that the myopic policy does not improve significantly on the random strategy due to the SU congestion while the proposed sensing policy benefits from multiuser diversity. In addition, by selecting strong channels, it boosts individual throughput. We also observe that the CSI-aided myopic approach does not completely eliminate collisions, so additional gains can be obtained by combining CSI-aware sensing with multiuser sensing strategies.

In Fig 4, we study the performance of the proposed and conventional sensing policies as a function of the average SNR $\mu_\gamma$. As expected, the performance improves for all sensing policies as $\mu_\gamma$ increases. Note also that the gain associated with CSI-aided strategies increases with average SNR.

B. *Correlated Lognormal Shadow Fading Model*

Next, we explore adaptation to the long-term fading for the same CR network. We employ the model [15], where the shadow fading of different links on each channel is correlated due to the proximity of SUs and the fading coefficient is fixed within $T = 20$ time slots and is the same for all channels for each SU pair. Denote the SNR (in dB) as $\gamma_{dB}$, which is Gaussian with mean $\mu_{\gamma_{dB}}$ and standard deviation $\sigma_{\gamma_{dB}}$, where

$$\mu_{\gamma_{dB}} = 10\log_{10}(\mu_\gamma) - \frac{\sigma_{\gamma_{dB}}^2}{2\xi} \text{ with } \xi = 10/\ln 10 \text{ and } \mu_\gamma = E[\gamma].$$

We denote the correlation coefficient between any two links as

$$\rho_{mm'} = \frac{E[(\gamma_{dB}^m - \mu_{\gamma_{dB}})(\gamma_{dB}^{m'} - \mu_{\gamma_{dB}})]}{\sqrt{E[(\gamma_{dB}^m - \mu_{\gamma_{dB}})^2]E[(\gamma_{dB}^{m'} - \mu_{\gamma_{dB}}^{m'})^2]}}.$$

We assume $\rho_{mm'} = \rho^{|m-m'|}$, i.e., the correlation among SU links decreases with index difference.

We compare performance of the proposed CSI-aided sensing policy and conventional strategies for $\mu_{\gamma_{dB}} = 10dB$, $\rho = 0.2$ and $\sigma_{\gamma_{dB}} = 5dB$ in Fig 5. We observe that the performance of the proposed CSI-aided policy is still much better than that of the myopic policy and the random selection policy.

The impact of different values of $\rho$ is shown in Fig 6. Note that the throughput degrades as $\rho$ increases. For small $\rho$, the CSI-aided myopic policy randomizes decisions and benefits from multiuser diversity However, when the SU links are highly correlated (for $\rho \approx 0.9$), the performance of the proposed CSI-aided sensing policy approaches that of the myopic policy. In this case, all SU links actually experience almost the same fading patterns, so the multiuser diversity gained from adapting to the channel conditions is lost.

Typically, the correlation of shadow fading is very small [16, 17]. Moreover, even if the correlation between two nearby links is high, it is seldom the case that many links in the neighborhood are highly correlated. So in general, the wireless environment can provide sufficient multiuser diversity to randomize the sensing decisions.

## IV. CONCLUSION AND FUTURE WORK

We proposed to incorporate CSI into the MAC protocol design of CR networks and demonstrated that this method improves the SU throughput and achieves multiuser diversity by randomizing the sensing decisions. . While the proposed policy reduces secondary collisions, it does not completely eliminate them. Therefore, we plan to combine CSI-aware sensing with a multiuser MAC approach to obtain additional gains.

Moreover, we assumed full knowledge of the CSI at the SU transmitter. However, this assumption is not applicable in practical mobile radio channels. We will investigate CSI estimation and prediction in CR networks for realistic channel models that include path loss, long-term and short-term fading, and interference. In addition, we modeled the PU traffic as a stationary Markov chain with known transition matrix. However, the transition matrix may vary over time, so we plan to consider scenarios where SUs need to learn and track the PU traffic statistics as well as channel characteristics.

Finally, we have employed the channel capacity to explore the potential of CSI-aided MAC in CR networks. One natural extension of this work is to investigate adaptive transmission methods that realize this potential in practice.

REFERENCE

[1] I.F. Akyildiz, W.Y. Lee, M.C. Vuran and S. Mohanty, "A survey on spectrum management in cognitive radio networks," IEEE Communications Magazine, vol. 46, pp. 40-48, 2008.

[2] J. Jia, Q. Zhang and X. Shen, "HC-MAC: A hardware-constrained cognitive MAC for efficient spectrum management," IEEE Journal on Selected Areas in Communications, vol. 26, pp. 106-117, 2008.

[3] Q. Zhao, L. Tong, A. Swami and Y. Chen, "Decentralized cognitive MAC for opportunistic spectrum access in ad hoc networks: A POMDP framework," IEEE Journal on Selected Areas in Communications, vol. 25, pp. 589-600, 2007.

[4] H. Liu and B. Krishnamachari, "Randomized strategies for multi-user multi-channel opportunity sensing," in IEEE CCNC Cognitive Radio Networks Workshop, 2008.

[5] K. Liu, Q. Zhao and Y. Chen, "Distributed sensing and access in cognitive radio networks," in Proc. of 10th International Symposium on Spread Spectrum Techniques and Applications (ISSSTA), 2008.

[6] L. Lai, H. El Gamal, H. Jiang and H.V. Poor, "Cognitive medium access: Exploration, exploitation, and competition," IEEE Transactions on Mobile Computing, pp. 239-253, 2010.

[7] H. Liu, B. Krishnamachari and Q. Zhao, "Cooperation and learning in multiuser opportunistic spectrum access," in IEEE International Conference on Communications Workshops (ICC '08), pp. 487-492, 2008.

[8] H. Su and X. Zhang, "Cross-layer based opportunistic MAC protocols for QoS provisionings over cognitive radio wireless networks," IEEE Journal on Selected Areas in Communications, vol. 26, pp. 118-129, 2008.

[9] Q. Zhao and B. Krishnamachari, "Structure and optimality of myopic sensing for opportunistic spectrum access," in IEEE International Conference on Communications (ICC'07), pp. 6476-6481, 2007.

[10] A. Duel-Hallen, "Fading channel prediction for mobile radio adaptive transmission systems," Proc IEEE, vol. 95, pp. 2299-2313, 2007.

[11] J. So and N.H. Vaidya, "Multi-channel mac for ad hoc networks: handling multi-channel hidden terminals using a single transceiver," in Proceedings of the 5th ACM international symposium on Mobile ad hoc networking and computing, pp. 222-233, 2004.

[12] J. Chen, S.T. Sheu and C.A. Yang, "A new multichannel access protocol for IEEE 802.11 ad hoc wireless LANs," in 14th IEEE Proceedings on Personal, Indoor and Mobile Radio Communications (PIMRC'03), pp. 2291-2296, 2003.

[13] A.J. Goldsmith and S.G. Chua, "Variable-rate variable-power MQAM for fading channels," IEEE Transactions on Communications, vol. 45, pp. 1218-1230, 1997.

[14] N. Mehta, A. Duel-Hallen and H. Hallen, "Template design and propagation gain for multipath UWB channels with per-path frequency-dependent distortion," in IEEE Military Communications Conference (MILCOM'09), pp. 1-7, 2009.

[15] T. Klingenbrunn and P. Mogensen, "Modelling cross-correlated shadowing in network simulations," in IEEE VTS 50th Vehicular Technology Conference (VTC'99), pp. 1407-1411, 1999.

[16] P. Agrawal and N. Patwari, "Correlated link shadow fading in multi-hop wireless networks," IEEE Transactions on Wireless Communications, vol. 8, pp. 4024-4036, 2009.

[17] A. Hong, M. Narandzic, C. Schneider and R.S. Thoma, "Estimation of the correlation properties of large scale parameters from measurement data," in IEEE 18th International Symposium on Personal, Indoor and Mobile Radio Communications (PIMRC'07), pp. 1-5, 2007.